\documentclass[runningheads]{llncs}

\usepackage[T1]{fontenc}
\usepackage{graphicx}

\usepackage{comment}
\usepackage{amsmath}

\begin{document}
	
	\title{Evaluating the Impact of Nonlinear Registration on Longitudinal PET Analysis in a Rodent Stroke Model}
	
	\author{Haytham Ziani\inst{1} \and El Mostafa Kalmoun\inst{1} \and Zohaib Atif\inst{2} \and Hyuk-Sang Kwon\inst{2}}
	
	\titlerunning{Nonlinear PET Registration in a Rodent Stroke Model}
	\authorrunning{H.Ziani et al.}
	
	\institute{School of Science and Engineering, Al Akhawayn University in Ifrane, Morocco \email{h.ziani@aui.ma} \and
		3D-BIT Lab, Department of Biomedical Science and Engineering, Gwangju Institute of Science
		and Technology, Gwangju 61005, South Korea}
	
	\maketitle
	\begin{abstract}
		Longitudinal rodent positron emission tomography (PET) studies of stroke and neuromodulation rely on accurate registration to a common atlas space to detect group-level effects. Prior work applying ultra-high frequency low-intensity magnetic stimulation (UHF-LiMS) to a rat model of chronic capsule infarct used affine normalization to a Sprague Dawley MRI template without evaluating that choice \cite{atif2026}. We compare a two-stage (affine and nonlinear) pipeline against an affine-only one for rat PET-to-atlas alignment on the same dataset. Across 32 longitudinal $[^{18}\mathrm{F}]$-FDG PET scans (8 rats, 4 timepoints), the nonlinear stage gave a higher Dice similarity coefficient on every scan, and the gain was significant at the rat level. Jacobian determinants showed no folding on the sampling grid. Group-level linear mixed-effects maps of change from baseline (AFNI 3dLME), computed identically under both pipelines, differ in the size, number and location of their suprathreshold clusters. In this dataset, the registration pipeline is therefore a variable that can change the interpretation of downstream group-level analyses.

	\end{abstract}

	\section{Introduction}
	Longitudinal small-animal positron emission tomography (PET) imaging is a standard tool for tracking disease progression and treatment response over time. A critical step in such studies is registration: aligning each scan, often acquired weeks apart under variable noise and motion conditions, into a common atlas space. The spatial accuracy of this step directly affects the reliability of subsequent group-level statistics across subjects and timepoints.

	In rodent PET studies, registration is often treated as a routine step rather than as a methodological variable \cite{pascau2008}. A single affine transformation to a template is commonly applied and followed directly by group-level analysis, without quantifying registration accuracy or comparing alternative alignment strategies \cite{vallez2015}. Given the small size of rodent brains, their susceptibility to motion and the limited resolution of small-animal PET, residual misalignment after affine-only registration can bias quantitative outcomes \cite{pascau2008,nandi2024}. The study that introduced ultra-high-frequency, low-intensity magnetic stimulation (UHF-LiMS) for chronic capsular infarct in rats \cite{atif2026} illustrates this gap. Its longitudinal PET scans were normalized to a Sprague Dawley magnetic resonance imaging (MRI) template by affine registration alone, and no registration accuracy was reported.

	On that same dataset, we evaluate a two-stage pipeline: affine registration followed by nonlinear Symmetric Normalization (SyN). Registration quality is assessed with the Dice similarity coefficient and Jacobian determinant maps. The downstream effect is assessed with group-level linear mixed-effects maps computed under both pipelines.

	\section{Related Work}
	Affine or rigid-body normalization to a species-standard atlas dominates small-animal PET. The practice comes from human neuroimaging, where higher resolution and lower inter-subject variability often make it adequate \cite{pascau2008}. The assumption is weaker in rodent stroke models, where focal lesions introduce local, subject-specific deformation that no single linear transform can capture \cite{pascau2008,vallez2015,nandi2024}. Deformable methods add a local, diffeomorphic correction on top of an initial affine alignment. SyN \cite{avants2008}, implemented in the Advanced Normalization Tools (ANTs) \cite{avants2009}, is among the most established of these methods and is the one adopted here.

	Learning-based methods such as VoxelMorph \cite{balakrishnan2019} and SynthMorph \cite{hoffmann2022} predict a deformation field with a network trained once, instead of solving a new optimization problem for each image pair. A single cohort of 32 PET scans registered to an MRI atlas is too small to train or adapt such a model, so both pipelines evaluated here are classical, optimization-based ones.

	\section{Methods}

	\subsection{Data and Preprocessing}

	We use the 5 mT UHF-LiMS dataset of Atif et al. \cite{atif2026}: longitudinal PET with $[^{18}\mathrm{F}]$\hspace{0pt}fluoro\-deoxy\-glucose (FDG) from 8 rats with photothrombotic capsular infarct, each imaged at four timepoints (32 scans). In the original study the images were attenuation-corrected and reconstructed by three-dimensional ordered-subsets expectation maximization (3D-OSEM). We denote the timepoints $T_0$ (baseline, pre-stroke), $T_1$ (2 weeks post-stroke, pre-stimulation), $T_2$ (1 week post-stimulation) and $T_3$ (2 weeks post-stimulation). They correspond to the condition labels \texttt{d00}, \texttt{d01}, \texttt{d07} and \texttt{d14} in the analysis scripts, where the suffix indexes the stimulation protocol rather than elapsed days.

	The inherited volumes are not raw reconstructions. In \cite{atif2026} they had already been resampled with \texttt{3dresample} from the Analysis of Functional NeuroImages (AFNI) suite \cite{afni1996} onto the grid of the Waxholm Space (WHS) Sprague Dawley MRI atlas \cite{whs2014} that serves as our working space ($52\times50\times52$ voxels of $0.388\times0.796\times0.388$~mm). They had also been scaled by a per-scan constant and affine-aligned to a study-specific mean template with \texttt{3dWarpDrive}. Both pipelines compared here therefore start from a linearly aligned state, a point we return to in Section~\ref{sec:discussion}. Residual translations and rotations that could impede convergence were then removed manually with AFNI's \texttt{nudge\_dataset} plugin under visual inspection.

	\subsection{Affine Registration}
	The pre-aligned scans were registered with a 12-degree-of-freedom affine transformation by AFNI's \texttt{align\_epi\_anat.py} (\texttt{epi2anat} mode). The target was the native-resolution WHS atlas ($200\times400\times200$ voxels at $0.1$~mm), and the result was written back onto the source grid. The cost function was normalized mutual information (NMI). Unlike metrics that assume a monotonic intensity relationship, NMI is robust to differing fields of view, low spatial resolution and metabolic contrast variation \cite{maes1997,pluim2003}. Every registration was checked visually.

	\subsection{Nonlinear Registration}
	The affinely registered scans were then refined by SyN in ANTsPy, the Python interface of ANTs \cite{ants2011}, against the atlas resampled to the working grid. We used the Mattes mutual information metric \cite{mattes2003} with 32 histogram bins and the deformation-only \texttt{SyNOnly} transform, so that this stage refines the affine result instead of re-estimating it. The linear component returned with the warp was the identity in all 32 cases, which confirms that the stage is purely deformable.

	Before group-level modeling, AFNI's \texttt{3dmerge} smoothed the registered volumes of both pipelines with the same Gaussian kernel of $1.2$~mm full width at half maximum (FWHM). Smoothing affects only the statistical maps, and the registration metrics of Section~\ref{sec:metrics} use the unsmoothed volumes. We report the kernel explicitly because smoothing directly influences cluster extent.

	\subsection{Evaluation Metrics}\label{sec:metrics}
	Let $M$ denote a moving PET volume, $F$ the fixed atlas, and $\phi$ the estimated transformation (affine, or affine followed by SyN), so that $M(\phi)$ is the registered volume. For binary masks $A$ and $B$, with $|\cdot|$ the number of voxels, the Dice similarity coefficient \cite{dice1945} is $\mathrm{Dice}(A,B) = 2|A\cap B|/(|A|+|B|)$. We take $B=S^F$, the binarized whole-brain atlas mask, which also restricts the group-level model and contains $N = 18{,}197$ voxels. Because the PET volumes carry no anatomical labels, we take $A=S^M(\phi)$, the set of the $N$ brightest voxels of $M(\phi)$. Both masks then have the same cardinality, so volume is not a confound and the score depends only on where the most metabolically active voxels fall relative to the atlas brain. The score is thus an intensity-derived proxy for the alignment of the high-uptake envelope, not a label-based overlap of delineated structures. We use it because no per-structure segmentations of these volumes exist, and we discuss its limits in Section~\ref{sec:discussion}.

	The determinant of the Jacobian matrix $J_\phi$ of each deformation served as a regularity check. A diffeomorphism has $\det J_\phi>0$ everywhere, and non-positive values indicate folding. We computed $\det J_\phi$ from the forward warp on the atlas grid, so the check is discrete and does not prove invertibility of the continuous map.

	Group-level effects were modeled voxelwise with AFNI's \texttt{3dLME} \cite{chen2013}, with timepoint as a fixed effect and rat as a random intercept, inside the mask $S^F$. The model was fitted separately to the affine-only and the Affine + SyN data, so the two pipelines share data, mask and contrasts ($T_1-T_0$, $T_2-T_0$, $T_3-T_0$). Maps of the voxelwise $Z$ statistic were thresholded at $p<0.001$ (two-sided $|Z|>3.29$) and clustered with AFNI's \texttt{3dClusterize} (face and edge neighbors, minimum extent 42 voxels, about $5.03\,\mu\mathrm{L}$). This extent was fixed a priori to suppress isolated voxels and was not derived from a cluster-size null distribution. The maps are therefore uncorrected, and they serve to compare topology between pipelines, not to establish significance.

	Scan-level Dice differences (SyN minus affine) are reported descriptively. Since the 32 scans are nested within 8 rats, the confirmatory test uses the 8 per-rat mean differences: a one-sample $t$-test against zero, with an exact binomial sign test. A linear mixed-effects model of the scan-level difference (timepoint fixed, rat as random intercept, Satterthwaite degrees of freedom) tests whether the gain varies across timepoints.

	\section{Results}

	\subsection{Registration Accuracy}
\begin{figure}[!htb]
		\centering
		\includegraphics[width=0.42\textwidth]{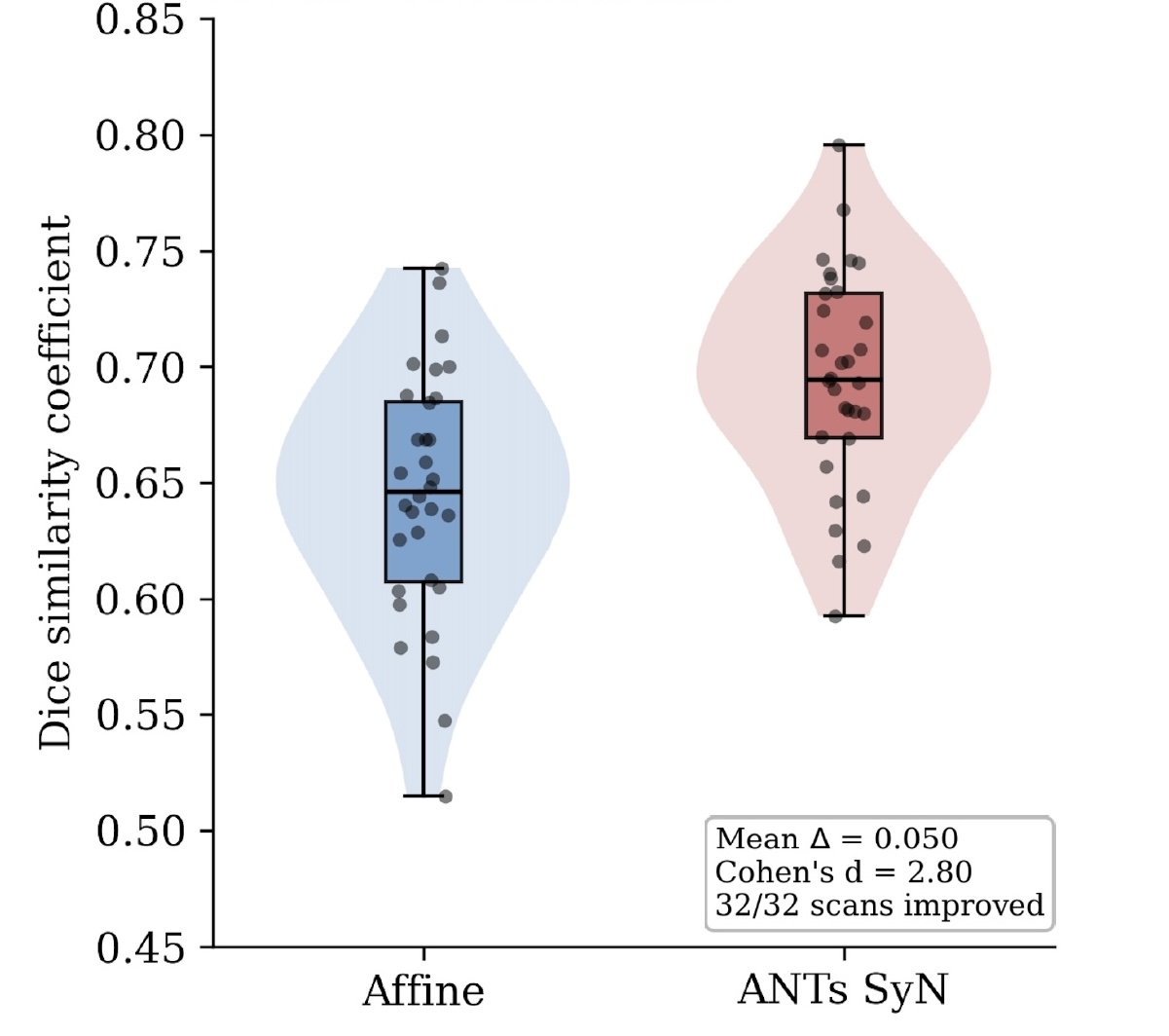}
		\caption{Scan-level Dice similarity coefficients for affine-only and Affine + SyN registration (32 rodent PET scans).}
		\label{fig:dice}
	\end{figure}

	The Affine + SyN pipeline gave a higher Dice than affine-only registration on all 32 scans (Figure~\ref{fig:dice}), with scan-level gains between $0.024$ and $0.101$. Mean Dice rose from $0.6449$ (standard deviation $\mathrm{SD}=0.0528$) to $0.6952$ ($\mathrm{SD}=0.0464$). The scan-level gain averaged $0.0504$ ($\mathrm{SD}=0.0180$), a large effect (Cohen's $d=2.80$). At the rat level the mean gain was $0.0504$ ($\mathrm{SD}=0.0103$), with $t(7)=13.84$, $p=2.4\times10^{-6}$ and 95\% confidence interval $[0.0417, 0.0590]$. All 8 rats improved (sign test, $p=0.0078$). The mixed-effects model detected no dependence of the gain on timepoint ($p=0.053$, $0.745$ and $0.729$ for $T_1$, $T_2$ and $T_3$ against $T_0$), although eight rats give this test little power and the $T_1$ value is borderline. Both pipelines lie far above chance: a randomly placed $N$-voxel mask has expected Dice $N/N_{\mathrm{FOV}} \approx 0.135$, where $N_{\mathrm{FOV}}=52\times50\times52$ is the number of voxels in the field of view.

	\subsection{Diffeomorphism Check}
	Over the 32 determinant maps, $\det J_\phi$ ranged from $0.200$ to $2.964$, with a per-map mean of $1.000$, and no non-positive value occurred at any grid point. The deformations therefore show no folding on the evaluated grid. The implied local volume change lies between a fivefold contraction and a threefold expansion.

	\subsection{Group-Level Statistical Maps}

\begin{figure}[t]
		\centering
		\includegraphics[width=\textwidth]{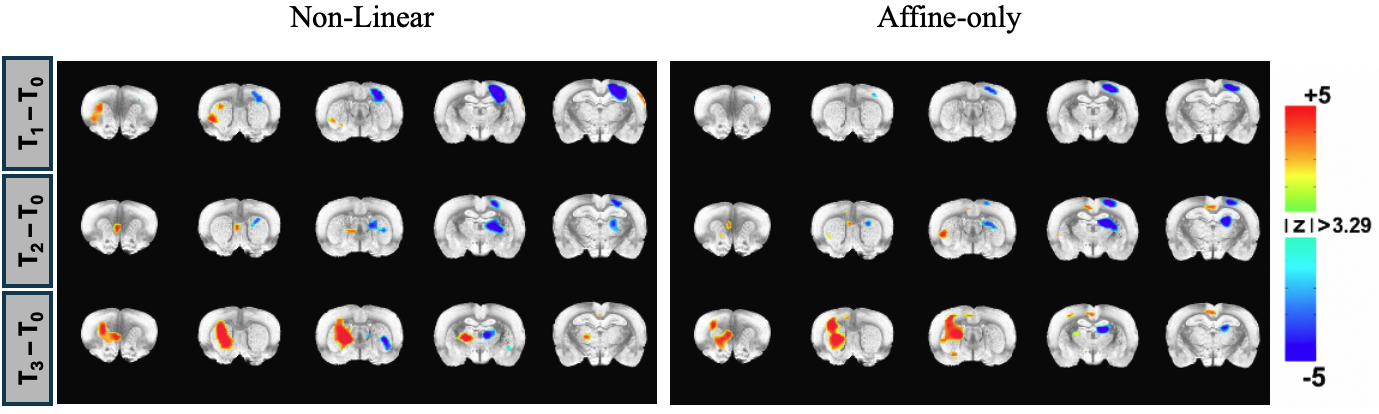}
		\caption{Thresholded group-level 3dLME maps for the nonlinear (Affine + SyN, left) and affine-only (right) pipelines. Rows are the contrasts $T_1-T_0$, $T_2-T_0$ and $T_3-T_0$. The overlay applies the voxelwise threshold ($p<0.001$, $|Z|>3.29$) and the 42-voxel extent threshold. The color scale saturates at $|Z| = 5$, and cool and warm colors denote negative and positive effects. Maps are uncorrected for multiple comparisons.}
		\label{fig:lme}
	\end{figure}

	Figure~\ref{fig:lme} shows the thresholded maps, and we write $Z_{\mathrm{peak}}$ for the extreme $Z$ value within a cluster. All contrasts are cumulative changes from the pre-stroke baseline $T_0$. They combine the effect of the infarct with that of any later stimulation and do not isolate a stimulation effect.

	Cluster topology depended markedly on the pipeline. At $T_1$, the largest affine-only cluster had 156 voxels ($Z_{\mathrm{peak}} = -9.61$) and the suprathreshold total was 474 voxels. With Affine + SyN, the largest cluster was about three times larger (476 voxels, $Z_{\mathrm{peak}} = -11.03$) and the total was 804 voxels. Both maps had four clusters at this timepoint.

	The pipelines were closest at $T_2$: 590 suprathreshold voxels in six clusters (largest 164, $Z_{\mathrm{peak}} = -8.25$) for affine-only, against 435 voxels in five clusters (largest 148, $Z_{\mathrm{peak}} = -5.60$) for Affine + SyN.

	At $T_3$ the divergence took another form. Affine + SyN gave more suprathreshold volume (938 vs.\ 806 voxels), but its largest cluster (437 voxels, $Z_{\mathrm{peak}} = 7.24$) was smaller than the affine-only one (593 voxels, $Z_{\mathrm{peak}} = 5.89$), and the rest of its signal was spread over seven smaller clusters. In this dataset, the pipeline thus changes the size, number and location of the suprathreshold clusters.

	\section{Discussion}\label{sec:discussion}
	The nonlinear stage improves overlap on every scan, and its Jacobian determinants stay well away from zero. The changes in cluster topology are therefore unlikely to come from folding or extreme local deformation. The volumes were already affinely aligned to a study mean before our pipeline began. We did not measure the gain of our affine stage separately, but this prior alignment suggests that most of the remaining correction comes from the SyN stage.

	At $T_1$, the larger primary cluster under Affine + SyN is consistent with reduced inter-subject misalignment of the lesion signal. At $T_3$, the spread of the signal over more numerous, smaller clusters may reflect finer localization of a spatially heterogeneous cumulative effect, or a change in noise sensitivity once residual misalignment is reduced. The present data cannot distinguish these two explanations.

	Two limitations bound these conclusions. First, with $n=8$ rats the asymptotic approximations behind \texttt{3dLME} are less reliable. The uncorrected maps therefore show how residual misalignment shifts signal consolidation, and they make no biological claim about the 5~mT stimulation effect. Second, the top-$N$ Dice measures agreement between the high-uptake envelope and the atlas brain. It is sensitive to gross position, scale and shape, but it cannot detect structures displaced within a correctly placed envelope, and it inherits any bias of the tracer distribution. The observed gain is a necessary but not sufficient condition for better anatomical correspondence. Establishing the latter would require label-based evaluation on segmented structures, which this dataset does not provide.

	\section{Conclusion}
	This work proposes no new registration algorithm. On a previously published UHF-LiMS stroke and stimulation dataset, adding a nonlinear stage to an affine-only pipeline significantly improved the overlap between the registered PET volumes and the atlas brain, without detectable folding. The same refinement changed the size, number and location of the clusters in the group-level 3dLME maps. In this dataset, registration was thus a methodological variable with a visible effect on the group-level results, and its evaluation, absent from the original study \cite{atif2026}, deserves a place in such analyses. Future work should cover other rodent disease models and compare classical and learning-based registration in small cohorts.

	\begin{credits}
		\subsubsection*{\ackname}
		This work was primarily conducted when the first author was visiting the 3D-BIT Lab within the Gwangju Institute of Science and Technology (GIST) under the Global Intern Program (GIP).

		\subsubsection{\discintname}
		The authors have no competing interests to declare that are relevant to
		the content of this article.
		\subsubsection*{Ethics declaration}
		The PET data were acquired in a previous study \cite{atif2026}, whose animal procedures were approved by the GIST Animal Care and Use Committee (approval no.\ GIST-2019-074) and reported following the Animal Research: Reporting of In Vivo Experiments (ARRIVE) guidelines. The present work is a secondary analysis of those images and involved no new animal procedures and no human participants.

		\subsubsection*{Data availability}
		The data analyzed in this study were acquired at the 3D-BIT Lab (GIST) for \cite{atif2026} and are used with the laboratory's permission.

	\end{credits}

\end{document}